\documentclass[pre,preprint,showpacs,preprintnumbers,amsmath,amssymb]{revtex4}
\usepackage{graphicx}% Include figure files

\usepackage{dcolumn}% Align table columns on decimal point
\usepackage{bm}% bold math
\usepackage[mathscr]{eucal}
\usepackage{mathrsfs}

\begin{document}

% Full title of the paper (Capitalized)
\title{Interpretation of the apparent activation energy of glass transition}
% \title{A thermodynamics investigation on glass transition}

% Authors, for the paper (add full first names)
\author{Koun Shirai}
% \email{koun@sanken.osaka-u.ac.jp}
\affiliation{%
The Institute of Scientific and Industrial Research, Osaka University, 8-1 Mihogaoka, Ibaraki, Osaka 567-0047, Japan
}%

% Abstract (Do not use inserted blank lines, i.e. \\) 
\begin{abstract}
The temperature dependence of the viscosity of glass is a major concern in the field of glass research. Strong deviations from the Arrhenius law make the interpretation of the activation energy difficult. In the present study, a reasonable interpretation of the apparent activation energy is demonstrated along similar lines as those adopted in solid-state physics and chemistry. 
In contrast to the widely held view that phase transition in glass occurs at the reference temperature $T_{0}$ according to the Vogel--Fulcher--Tammann formula, in the present work the transition observed at the glass-transition temperature $T_{g}$ is regarded as a phase transition from the liquid to solid phases.
A distinct feature of glass is that the energy barrier significantly changes in the transition range with width $\Delta T_{g}$. This change in the energy barrier alters the manner in which the apparent activation energy constitutes the Arrhenius form. 
Analysis of available experimental data showed that the actual energy barrier is significantly smaller than the apparent activation energy, and importantly, the values obtained were in the reasonable range of energy expected for chemical bonds. The overestimation of the apparent activation energy depends on the ratio $T_{g}/\Delta T_{g}$, which explains the existence of two types of glasses strong and fragile glasses. 
The fragility can be re-interpreted as an indication of the degree of increase in the energy barrier when approaching $T_{g}$ from high temperatures.
Since no divergence in viscosity was observed below $T_{g}$, it is unlikely that a transition occurs at $T_{0}$.
% The principle underlying the above conclusion is rigorous definition of state variables associated with the definition of equilibrium in thermodynamics theory.
\end{abstract}

% deflect from

\maketitle
%%%%%%%%%%%%%%%%%%%%%%%%%%%%%%%%%%%%%%%%%%
%% Only for the journal Gels: Please place the Experimental Section after the Conclusions

%%%%%%%%%%%%%%%%%%%%%%%%%%%%%%%%%%%%%%%%%%

\section{Introduction}
\label{sec:intro}

% Davies53a, Jackle86,Biroli13
The core problem in glass research at present is what is the nature of the glass transition \cite{Angell00,Lubchenko07,Berthier11,Stillinger13}. 
Grass transition is observed, for example, from specific heat ($C_{p}$) versus temperature ($T$) curves. The transition temperature $T_{g}$ is the temperature at which the specific heat exhibits a quick change. It can be experimentally observed, but its value varies to some extent depending on experimental conditions such as the cooling rate. This gives rise to the believe that $T_{g}$ is not an intrinsic property of glass.
% Hence, there is no consensus on whether it is as an intrinsic property of glass.
Unlike the usual phase transitions, structural relaxation plays a crucial role in glass transition: the status of experiments on relaxation is reported in review papers \cite{Ediger17,Niss18,Royall18}. The structural relaxation is reflected by viscosity $\eta$.
The viscosity of glass-forming liquid increases drastically by more than ten orders of magnitude when the liquid temperature approaches $T_{g}$ from a higher temperature (Fig.~\ref{fig:Glass-trans} (a)). The $T$ dependence of viscosity is most commonly expressed by the Vogel--Fulcher--Tammann (VFT) formula \cite{Vogel21,Fulcher25,Tammann26},
\begin{equation}
\eta(T) = C \exp \left( \frac{D}{T-T_{0}} \right),
\label{eq:VFT}
\end{equation}
where $C$, $D$, and $T_{0}$ are material-dependent constants. The reference temperature $T_{0}$ is less than $T_{g}$ and is generally $0.6$ to $0.8$ times $T_{g}$ \cite{Angell97}. While $T_{0}$ is obtained by extrapolation, many researchers have attempted to identify its significance as the genuine transition temperature, which is an intrinsic property of a specific glass. The relation of $T_{0}$ with the so-called Kauzmann temperature $T_{K}$ is being actively debated \cite{Ngai99,Ito99,Martinez01,Sastry01,Tanaka03,Wang06,Hecksher08}.
For normal liquids, the nature of viscosity is well understood as a thermally activated process, which obeys the Arrhenius law,
\begin{equation}
\eta(T) = \eta_{0} \exp \left( \frac{Q_{a}}{k_{\rm B} T} \right),
\label{eq:Act-eta}
\end{equation}
where $k_{\rm B}$ is Boltzmann's constant, and $\eta_{0}$ is a material-dependent constant. In this case, $Q_{a}$ is clearly the activation energy. This formula was derived from first principles \cite{Rate-Theory-Eyring}. For normal liquids, the $Q_{a}$ values are of the order of a few tenths of electron volts or less \cite{Ewell37}, which is a reasonable range considering the chemical energies of materials. % Ree55
If this standard formula, i.e., Eq.~({\ref{eq:Act-eta}}) is applied to the viscosity of glasses in a narrow range of temperature, $Q_{a}$ values are found to be surprisingly large more than 1 eV (in some cases, 10 eV) even for organic glasses whose melting temperature $T_{m}$ is less than room temperature \cite{Davies53,Davies53a,Hodge94}.  
There is no convincing theory to interpret the large values of $Q_{a}$. Furthermore, the pre-exponential factor $A= 1/\eta_{0}$ is extremely large: typically, the magnitude is of the order of $e^{100}$ \cite{Hodge94}. Thus far, no theory that accounts for this extraordinarily large value has been established.

At present, the meaning of activation energy of glass is not being actively researched. Instead, by normalizing $T$ with $T_{g}$ in the Arrhenius plot, which is now known as the Angell plot, researchers are attempting to discover something universal for glass transition \cite{Nemilov-VitreousState,Laughlin72, Angell76,Angell88,note-Nussinov}. 
In the normalized form of the Angell plot, how largely the $T$ dependence of $\eta$ deviates from the Arrhenius law is the main information that can be obtained; fragility is a parameter used to quantify the degree of the deviation. Depending on the fragility, glasses are classified as strong or fragile. 
Although this manner of plot is useful for classifying glasses, it does not aid the chemical interpretation of the activation energy, which is the basic element for understanding properties of materials.
Without a pertinent interpretation of the activation energy, one cannot understand why the viscosity of glass changes so largely when approaching $T_{g}$. It is necessary to appropriately interpret $Q_{a}$ obtained from the Arrhenius plot: this is the aim of this study.
% urges

A resolution of the problem of the activation energy of glass was achieved in a very different context, namely, in terms of the state variables of glass. Today, the standard view is that glass states are nonequilibrium states. 
Although the glass transition is a nonequilibrium phenomenon---transition itself represents nonequilibrium---unduly emphasizing the nonequilibrium character hides the thermodynamic nature of glass.
A frequently-claimed reason for the nonequilibrium character is that the state of glass is not determined by temperature and pressure ($p$) alone but is affected by the previous history of treatments. Thermodynamics states are to be specified solely by the current values of state variables; if the glass states were equilibrium, the properties must be completely determined solely by $T$ and $p$, but this is not true, from which the above conclusion is derived. 
% Hence, only conclusion that can be derived from this observation is that glass is not in equilibrium.
However, the attribution of this discrepancy to nonequilibrium state is an easy escape from the difficulty. Careful observation reveals that this dependence of the past treatment that the solid underwent is a common property of solids: any solid has, to a certain degree, the dependence of its properties on the preparation conditions. 
The fundamental question then is that besides $T$ and $p$ which could be the other state variables of a solid: the term {\it thermodynamic coordinate} (TC) is used to refer to state variable. 
This question was at times considered by several authors to be a flaw of the theory \cite{Bridgman50, Kestin70, Maugin94} but was not solved until recently. The author examined this topic by reappraising the definition of equilibrium and TCs for solids, and defined in a consistent manner. 
% Kestin92,Rice75
In short, the answer to the question of TCs for a solid is the time-averaged positions $\bar{\bf R}_{j}$ of all the constituent atoms of the solid: note that instantaneous positions are irrelevant, and only equilibrium positions have sense \cite{StateVariable,Shirai20-GlassState}. 
This consideration can also be true for the state during glass transition, provided some restrictions are imposed.
The thermal response of solids is so fast---typically 10 ps---that the fast vibrational motion is adiabatically decoupled from slow changes in the structure. This behavior is expressed as the adiabatic approximation of the second kind \cite{Shirai20-GlassState}. Then, the instantaneous state at timescales much longer than this response time can be regarded as an equilibrium state, i.e., {\it temporal equilibrium} \cite{Shirai20-GlassState}. The states in glass transition are states of temporal equilibrium. Even during such a transition, all the averaged positions $\bar{\bf R}_{j}(t)$ as well as the time-dependent temperature $T(t)$ are well defined, and accordingly, the glass state is expressed by the instantaneous values of all TCs. 

A striking consequence of this conclusion is that energy barrier $E_{b}$ of the structural relaxation of glass during glass transition is determined solely by the present positions of atoms, $E_{b} = E_{b}( \left\{ \bar{\bf R}_{j} \right\} )$, irrespective of the previous history. Since during the transition, the structure is mainly determined by temperature, the structural dependence of $E_{b}( \left\{ {\bf R}_{j} \right\} )$ can be represented as the temperature dependence of energy barrier $E_{b}(T)$. The Arrhenius analysis presumes that $Q_{a}$ is constant: however, this assumption does not hold for glass transition. Hence, the apparent activation energies obtained from the Arrhenius plot results in extraordinarily large values in the usual sense of the term ``chemical bond" \cite{Shirai20-GlassHysteresis}.

In this paper, the above conclusion regarding the temperature dependence of $E_{b}(T)$ is validated by analyzing experimental data on viscosity. Further, a pertinent interpretation of the apparent activation energy is provided. In a previous study \cite{Shirai20-GlassHysteresis}, this interpretation was derived from an analysis of the hysteresis in the $C$-$T$ curve. Viscosity measurement is a more direct method to obtain the activation energy. 
The rest of this paper is organized as follows: the theoretical basis for the present analyses is given in Sec.~\ref{sec:Backgrounds}. Then, Sec.~\ref{sec:Width} shows the first evidence for the proposed interpretation based on the relationship between the width of glass transition and the apparent activation energy. The second evidence is obtained by directly evaluating the temperature-dependent energy barrier: this evidence is described in Sec.~\ref{sec:TdependViscosity}, which provides three examples of classes of glasses. 
An important implication of the present results to glass transition is discussed in Sec.~\ref{sec:discussion}. Section \ref{sec:conclusion} concludes the study. 
Throughout this paper, the numerical values of viscosity $\eta$ are presented in terms $\log \eta$(poise) with base of 10.
%%%%%%%%%%%%%%%%%%%%%%%%%%%%%%%%%%%%%%%%%%%%%%%%

\section{Theoretical backgrounds}
\label{sec:Backgrounds}

\subsection{Glass state and glass transition}
\label{sec:terminology}
% \paragraph{Glass state.}
There exists a large gap in comprehension of the glass state between glass physics and other areas of solid-state physics. For non-experts in the field of glass research, glass is almost evidently a solid. The glass transition that occurs at $T_{g}$ is a phase transition between the solid and liquid phases. However, glass researchers do not think so. Many of glass researchers consider glass as a special type of liquid, specifically, {\it kinetically frozen liquid}. Hence, it is necessary at the outset to state the present view and to explain the differences from the viewpoint of glass literature.

\begin{figure}[htbp]
\centering
\includegraphics[width=95 mm, bb=0 0 560 760]{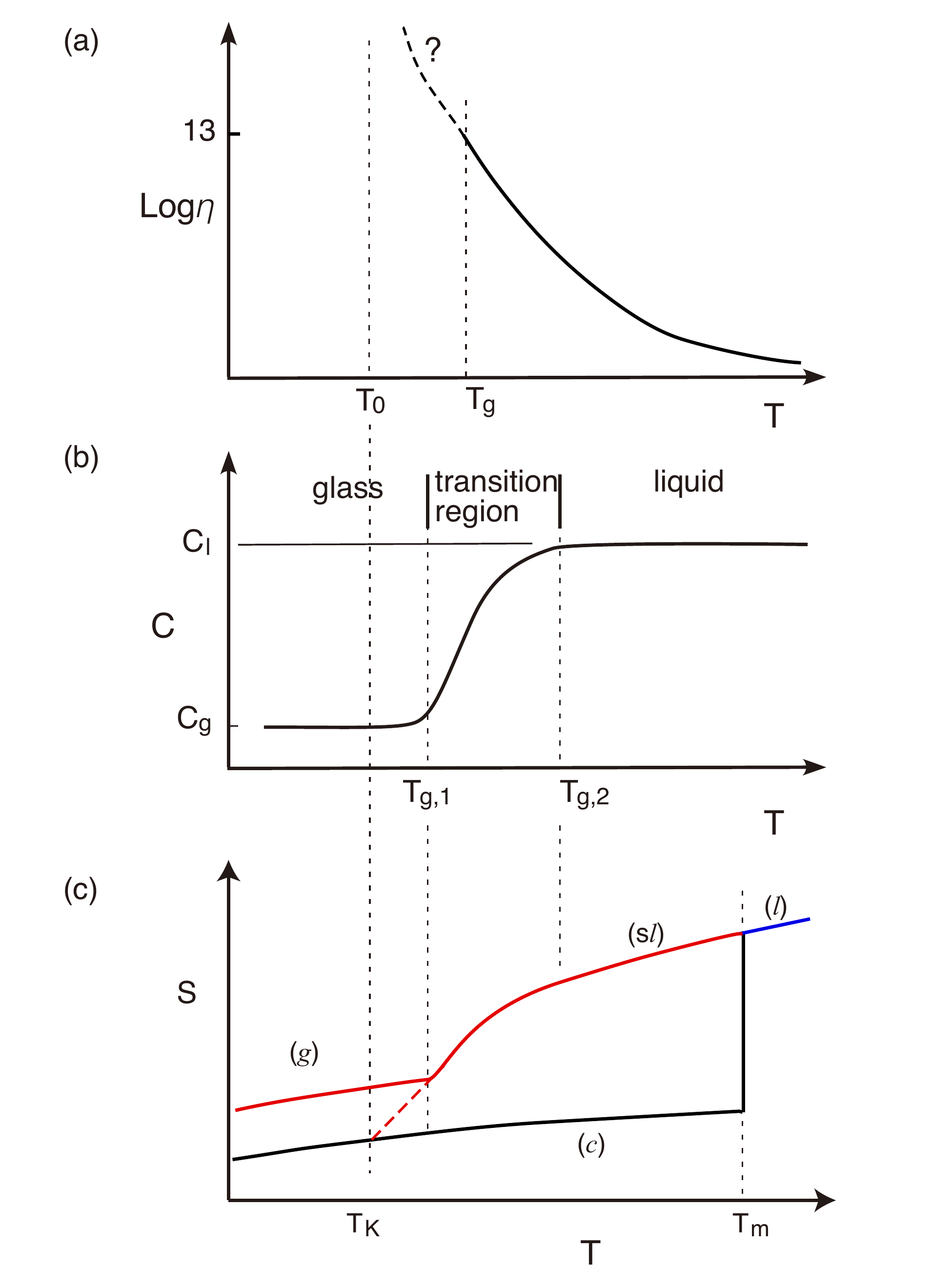} 
\caption{Glass transition represented in terms of viscosity $\eta$ (a) and specific heat $C_{p}$ (b). Evolution of the energy barrier for atomic movement is shown in (c).
} \label{fig:Glass-trans}
\end{figure}

The signature of the glass transition is observed in a quick change in the properties of glass, such as specific heat and thermal expansion, in a narrow range of temperature \cite{Kauzmann48}. Here, the range is specified by $T_{g,1}$ and $T_{g,2}$ with its width $\Delta T_{g} = T_{g,2}-T_{g,1}$, as shown in Fig.~{\ref{fig:Glass-trans}}(b).
In this paper, the term the {\it glass state} is restricted to refer to the phase of a glass substance below $T_{g,1}$, and is regarded as a solid phase. Above $T_{g,2}$ and below $T_{m}$, the state of the glass substance is a supercooled-liquid state; there $C_{p}$ obeys almost the classical limit of Dulong-Petitte law, $3R$ ($R$ is the gas constant) \cite{Angell95}. The state between $T_{g,1}$ and $T_{g,2}$ is referred to as {\it the transition state}. The $C_{p}$-$T$ curve often exhibits complicated shapes depending on the preparation conditions, which makes it difficult to identify the width $\Delta T_{g}$. Usually, an averaging value is used for $T_{g}$, unless special interest is paid for the width. In spite of the preparation dependence, it is meaningful to determine the unique value for $T_{g}$ within a reasonable range, provided suitable conditions are used \cite{Mazurin07,Moynihan76}.
In the relationship between the viscosity $\eta$ and $T$, in contrast, there is no characteristic temperature other than $T_{0}$ when $\eta(T)$ is recasted to the VFT form. However, it will be shown later that this is not true: see Sec.~\ref{sec:silicate}. Operationally, the $T_{g}$ can be defined as the temperature at which $\log \eta = 13$. % For correlation or the transition temperatures determined by the two methods, refer to \cite{Mazurin07}.
% (For practical purposes, further characteristic temperatures, namely, the strain point, annealing point, and softening points are used \cite{Lillie31}. The relation to $T_{g,1}$ and $T_{g,2}$ in the $C_{p}$ is, however, unclear.)
Many of glass researchers do not think that there is a sharp boundary between the supercooled-liquid state and the glass state. 
Even in the glass state, atoms are migrating with extremely slow velocities. From this view, $T_{g}$ cannot be the temperature of a phase transition. Instead, $T_{0}$ in Eq.~(\ref{eq:VFT}) deserves the genuine transition temperature, below which the so-called {\it ideal glass} appears. The reason that the ideal glass has not been observed is attributed to its extremely slow motions. From this, that the observed glass is thermodynamically a nonequilibrium state is concluded. 

The causes behind the above contrasting views are deeply rooted in the fundamental problems of thermodynamics: the definition of equilibrium and TC. The rigorous definition of equilibrium has been only recently established \cite{Gyftopoulos}, and it enables us to define TC in a consistent manner.  The author applied this principle to solids and found that the equilibrium positions of all the atoms $\left\{ \bar{\bf R}_{j} \right\}$ are TCs for solids \cite{StateVariable}. The present viewpoints are based on this quite general conclusion \cite{Shirai20-GlassState}. The important points relevant to this study are listed below, with contrasting the general views in glass research \cite{Ediger00,Lubchenko07,Chandler10,Berthier11,Biroli13,Stillinger13}: however, some of the views are not necessarily unique even in the glass research. % among glass researchers.

\begin{enumerate}
\item{{\em Glass is a solid}: it is viewed as a frozen liquid in glass literature.}
An often-claimed reason for the lack of distinction between the liquid and glass states is that atomic positions are random in both states. However, this is only true when the instantaneous positions of atoms $ {\bf R}_{j}(t) $ are compared. Thermodynamically, equilibrium positions $ \bar{\bf R}_{j} $, which are obtained by time averaging, have sense. They are well defined for solids, but not for liquids.

\item{{\em Glass is an equilibrium state}: it is viewed as a nonequilibrium state in glass literature.}
This is the natural consequence of the preceding conclusion. Although the changing state in the course of time during the transition is a nonequilibrium state, the values of TCs can be specified for each moment. When the change in the external conditions is halted and keep it constant for a time longer than the relaxation time, the transition state finally reaches the equilibrium state even in the transition region of temperature.

\item{{\em Glass transition that occurs at $T_{g}$ is a genuine phase transition}: it is viewed as an extrinsic change in glass literature.} Although the transition has dependence on the preparation conditions, this dependence, small or large, is a common feature of all phase transitions. In molecular-beam-epitaxy method, the growth temperature varies with the kinetics of source gases. 

\item{{\em The transition state is a mixture of liquid and solid phases}: it is viewed as a supercooled-liquid state after a sufficiently long time in glass literature.} Although the final states according to the two viewpoints are different, both share the common thought that the transition region is a {\it crossover} region between different phases. The term {\it dynamic heterogeneity} in glass literature is also compatible with the present view of the mixed state. 

\end{enumerate}

A natural consequence of the above-described views is that the energy barrier $E_{b}$ for atomic movement depends on the structure, that is, $E_{b}= E_{b}(\left\{ {\bf R}_{j} \right\})$, and hence it depends on temperature \cite{Shirai20-GlassHysteresis}. It follows that $E_{b}$ varies strongly in the transition region from the energy barrier of the solid state ($E_{b,g}$) to that of the liquid state ($E_{b,l}$), as shown schematically in Fig.~\ref{fig:Glass-trans}(c). 
The temperature dependence of the activation energy becomes clearer in molecular-dynamic simulations \cite{Debenedetti01,Han20}.
The idea of the $T$-dependent energy barrier was at times proposed by observing the deviations from the Arrhenius law \cite{Ward37,Gutmann52,Doolittle-two,Hildebrand76}. However, this idea did not gain wide acceptance.
% Notwithstanding, researchers did not consider this behavior of $E_{b}$ to be associated with the activation energy $Q_{a}$ in the Arrhenius plot for the glass transition for some reasons. 
The difficulty in interpreting the $T$ dependence of the viscosity---even if the structural dependence of energy barrier is accepted---is that the values of the activation energy at $T_{g}$ are too large on account of the magnitude of the energies of chemical bonds, as mentioned in Introduction. Further, the pre-exponential factor is too large: the absolute values of $\log \eta_{0}$ are greater than 100 for organic glasses \cite{Hodge94}.

% Angell plot \cite{Ito99,Ngai99,Martinez01,Sastry01,Tanaka03,Wang06,Hecksher08}

\subsection{Theory of viscosity}
\label{sec:TheoryOfViscosity}

Many theories have been proposed for explaining the viscosity of glass since Eyring derived the quantum-mechanical formula for viscosity of liquids \cite{Rate-Theory-Eyring,Eyring36}. These theories have been reviewed in textbooks by Nemilov \cite{Nemilov-VitreousState} and Rao \cite{Rao02}. The current trend of study has shifted to microscopic dynamics, focusing on topics such as the mode-coupling theory, dynamic facilitation, and first-order random transition \cite{Jackle86, Angell00,Ediger00,Sciortino05,Lubchenko07,Berthier11,Stillinger13,Biroli13}.
The approach employed in this study is the traditional one based on the reaction rate theory, with the aim of facilitating the chemical interpretation of the viscosity of glass.

For normal liquids, formula Eq.~(\ref{eq:Act-eta}), including the pre-exponential factor $\eta_{0}$, can be deduced from the microscopic theory, without assuming any model \cite{Rate-Theory-Eyring,Eyring36}. Eyring's theory explains well the viscosity of normal liquids \cite{Ewell37}. % Ree55
For an isotropic media, the pre-exponential factor $\eta_{0}$ in Eq.~(\ref{eq:Act-eta}) is expressed as
\begin{equation}
\eta_{0} = \left( \frac{a_{0}}{\lambda} \right)^{2} \frac{h}{a_{0}^{3}} \frac{Z_{n}}{Z_{a}},
\label{eq:A-factor}
\end{equation}
where $a_{0}$ is the mean interatomic distance, and $\lambda$ is the average distance between the equilibrium positions of the original and slipped states.
The product $(\lambda/a_{0})^{2} a_{0}^{3}$ may be interpreted as the activation volume $V_{a}$, which is the volume swept by the slipped atoms. Planck's constant $h$ is a universal constant but is equal to $k_{\rm B}T/\nu_{m}$ at thermodynamic equilibrium, where $\nu_{m}$ is the mean frequency of the frequency spectrum at $T$ and is interpreted as the attempt frequency. $Z_{n}$ and $Z_{a}$ are the partition functions of the normal and the activated states of the slipped atoms, respectively. Although it is difficult to calculate the partition function $Z$, the values of $Z$ in Eq.~(\ref{eq:A-factor}) are obtained only as the ratio $Z_{n}/Z_{a}$, which is of the order of unity in most cases, and hence, we can ignore this term.
By assuming $\lambda \approx a_{0}$, $\eta_{0} = h/V_{0}$ gives $\log \eta_{0} = -3.5$ for $V_{0} = 10\ {\rm \AA}^{3}$. This result is well in agreement with the experimental results: in many cases, $\log \eta_{0} = -3$ to $-4$ \cite{Barrer43}. The activation energy $Q_{a}$ is of the order of 0.1 eV or less \cite{Ewell37,Barrer43}.

% Schuh07
For solids, the definition of viscosity is not so simple because of its nonlinearity. Nevertheless, the rate theory of Eyring was widely applied to solid dynamics beyond the original object of viscosity. Plastic deformation of crystalline solids is explained by the rate theory, in which the motion is mediated by dislocations \cite{Schock80}. For the case of metallic glass, the motion is mediated by free volumes \cite{Spaepen77} or by shear transformation zones \cite{Argon77}. The physics involved in the processes is different from those of liquids. 
Despite this, the formal expression for the rate of deformation is almost the same as Eq.~(\ref{eq:A-factor}), if the quantities in this equation are suitably interpreted. 
Hence, there is no reason not to apply this formula for glasses in general. However, the problem in this case is that the material-dependent ``constants" are not constant during the transition, because the structure changes during the transition.

Now, let us ask the question why the $T$ dependence of the viscosity does not deviate from the Arrhenius law for strong glass. %contrary to the usual question in glass literature. 
From the present viewpoint, the energy barrier strongly varies with temperature, and hence a large deviation from the Arrhenius law is expected for all glasses. The apparent energy barrier $Q_{a}^{*}$ is obtained from the derivative of $\ln \eta$ with respect to $1/T$,
\begin{equation}
Q_{a}^{*}= \frac{\partial \ln \eta}{\partial (1/T)}.
\label{eq:QinArrhenius} 
\end{equation}
We are so accustomed to this formula to obtain energy barriers in numerous applications that we seldom consider how the Arrhenius form is altered when the energy barrier has $T$ dependence.
Let us consider a simple case of the linear dependence of the energy barrier on temperature,
\begin{equation}
E_{b}(T)=E_{b,g}-b (T-T_{g,1}),
\label{eq:EbdepT} 
\end{equation}
where $b=(E_{b,g}-E_{b,l})/\Delta T_{g} $ is a constant. In the Arrhenius form, the linear term in $T$ is canceled by the denominator in the exponent of Eq.~(\ref{eq:Act-eta}). Thus, Eq.~(\ref{eq:Act-eta}) becomes
\begin{equation}
\eta(T) = \eta_{0} e^{-b/k_{\rm B}} \exp \left( \frac{Q_{a}^{*}}{k_{\rm B}T} \right),
\label{eq:relax-tau1}
\end{equation}
where $Q_{a}^{*} = E_{b,g} + b T_{g,1}$. Hence, $T$ dependence is not seen in $Q_{a}^{*}$. Instead, there exists a large separation between $Q_{a}^{*}$ and $E_{b}$. Since $\Delta T_{g}$ is small, $Q_{a}^{*}$ is approximated by 
\begin{equation}
Q_{a}^{*} \approx \frac{T_{g}}{\Delta T_{g}} ( E_{b,g} - E_{b,l}).
\label{eq:Approx-Qa} 
\end{equation}
The factor $k=T_{g}/\Delta T_{g}$ acts as a magnification factor for the barrier height.
Although the linear term disappears from Eq.~(\ref{eq:Approx-Qa}), its effect appears in the pre-exponential factor as $\eta_{0}' = \eta_{0} e^{-b/k_{\rm B}}$. 
Sometimes, the entropy of atomic migration contributes to the transport coefficients through the $T$-independent term $e^{b/k_{\rm B}}$ in a manner similar to Eq.~(\ref{eq:relax-tau1}). However, the above derivation shows that it is not appropriate to interpret the $e^{b/k_{\rm B}}$ term in terms of entropy. In fact, $b/k_{\rm B}$ often becomes as large as 100. It is impossible to explain this magnitude by entropy. Even for vaporization, the increase in entropy is at most of the order of 10 times $k_{\rm B}$.
To obtain the genuine energy barrier $E_{b}$, the term $e^{-b/k_{\rm B}}$ should be retained in the Arrhenius analysis. Thus, $E_{b}$ can be obtained by 
\begin{equation}
E_{b}(T) = k_{\rm B}T \ln \left( \frac{\eta(T)}{\eta_{0}} \right),
\label{eq:trueEb}
\end{equation}
provided that $\eta_{0}$ is independent of $T$.
Unfortunately, $\eta_{0}$ may not be expected to remain constant during phase transition because the structure changes and does the size $\lambda$ of the moving unit. In addition, the ratio $Z_{n}/Z_{a}$ may not be ignored because the moving units become more collective motions for solids.
This problem has to be solved.
% One has to manage this problem by some means.

%%%%%%%%%%%%%%%%%%%%%%%%%%%%%%%%%%%%%%%%%%%%%%
%%%%%%%%%%%%%%%%%%%%%%%%%%%%%%%%%%%%%%%%%%%%%%
%%%%%%%%%%%%%%%%%%%%%%%%%%%%%%%%%%%%%%%%%%%%%%
\section{Width of glass transition}
\label{sec:Width}

\begin{figure}[htbp]
\centering
\includegraphics[width=120 mm, bb=0 0 360 250]{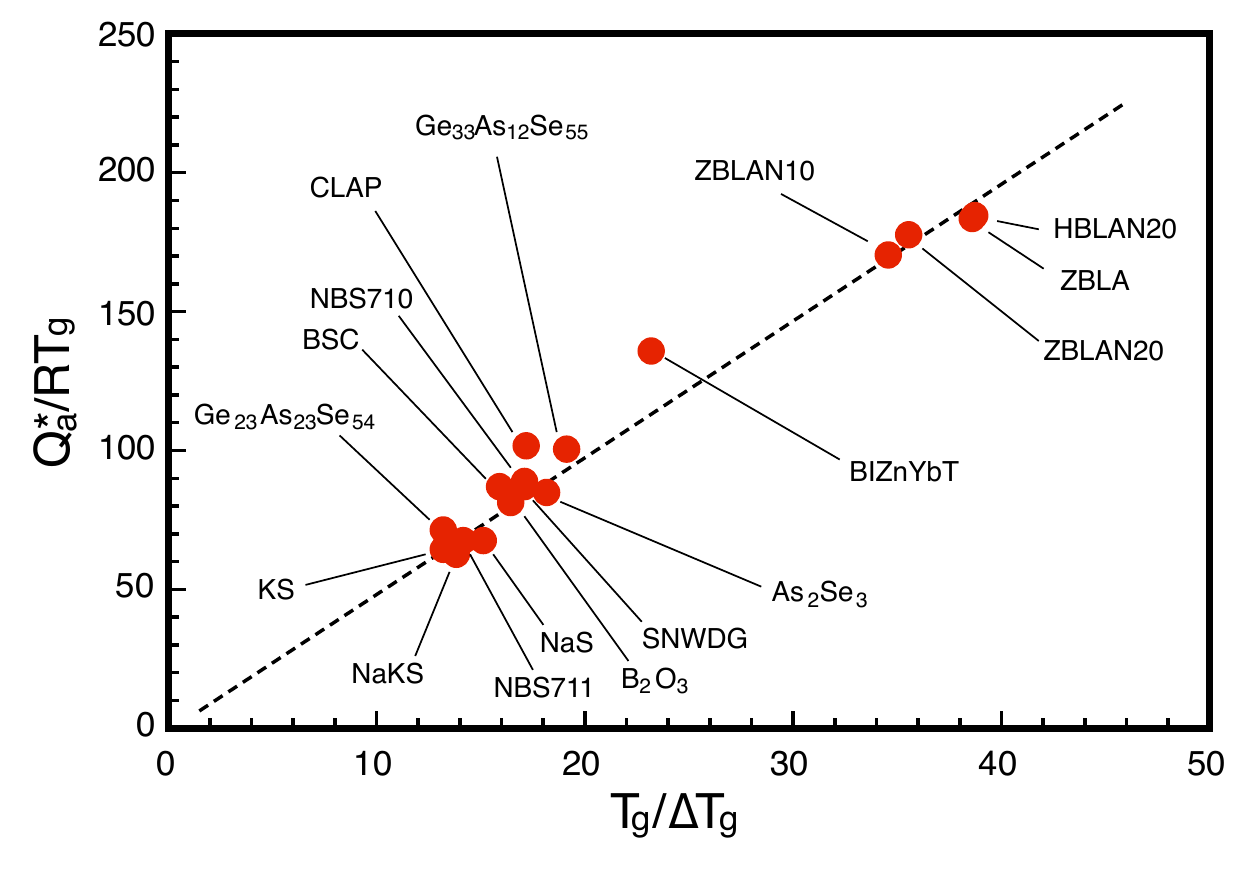} 
\caption{Plot of Moynihan's data showing the relationship between $T_{g}/\Delta T_{g}$ and $Q_{a}^{*}/k_{\rm B} T_{g}$ \cite{Moynihan95}. The sources of the original data are quoted in Ref.~\cite{Moynihan95}.
% The original data are: ${\rm As_{2}Se_{3}}$ \cite{Easteal77,Tatsumisago90}, ZBLAN20 \cite{note-Whang,Hasz92}, HBLAN20 \cite{note-Whang,Hasz92a}, ZBLAN10 \cite{note-Whang,Hasz92}, CLAP \cite{note-Whang2}, ${\rm B_{2}O_{3}}$ \cite{Moynihan76,DeBolt76,Macedo68}, ZBLA \cite{note-Whang,Hasz92}, ${\rm Ge_{23}As_{23}Se_{54}}$ \cite{Tatsumisago90}, BIZnYbT \cite{Bouaggad86,Christensen88}, ${\rm Ge_{33}As_{12}Se_{55}}$ \cite{note-Crichton,Webber76}, NBS711 \cite{Crichton88a,Napolitano74}, NaKS \cite{Moynihan76,Poole49}, NaS \cite{Moynihan76,Poole49}, KS \cite{Moynihan76,Poole49}, SNWDG \cite{Boulos80}, NBS710 \cite{Sasabe84,Napolitano74}, BSC \cite{Moynihan76,Macedo68}.
} \label{fig:Moynihan-plot}
\end{figure}

% \paragraph{Apparent activation energy.}
The present theory predicts that the apparent activation energy $Q_{a}^{*}$ is magnified by the factor $k=T_{g}/\Delta T_{g}$. Let us examine this magnification in $Q_{a}^{*}$ by analyzing the experimental data.
The data for this purpose are taken from measurements of the specific heat. While numerous data have been accumulated for $T_{g}$, data for $\Delta T_{g}$ are very rare. The width is sensitive to the shape of the $C$-$T$ curve in the transition region, which is largely affected by the conditions of sample preparation and measurement \cite{Moynihan76}.
It is, therefore, important to use the references in which the experimental conditions are well documented.
% obtained in an acceptable level of the difference in experimental conditions. 
Moynihan compiled data of $\Delta T_{g}$ for 17 inorganic glasses with $T_{g}$ well above room temperature \cite{Moynihan95}. These glasses include chalcogenides, heavy-metal fluorides, and network oxide glasses. 
He took special care when collecting experimental data: i.e., only the data obtained under common conditions, such as within an acceptable range of heating rate, were considered.
He found that there is a good correlation between $Q_{a}^{*}$ in viscosity measurement and $\Delta T_{g}$ in heat-capacity measurement
\begin{equation}
\frac{Q_{a}^{*}}{k_{\rm B} } \Delta \left( \frac{1}{T_{g}} \right) = 4.8 \ \pm 0.4.
\label{eq:MoynihanFormula} 
\end{equation}
The values of $Q_{a}^{*}$ are obtained in the range $11 \le \log \eta(T_{g}) \le 12$, where the $T$ dependence of viscosity obeys the Arrhenius law. 
The relationship between $T_{g}/\Delta T_{g}$ and $Q_{a}^{*}/k_{\rm B} T_{g}$ which is taken from Moynihan's data is plotted in Fig.~\ref{fig:Moynihan-plot}. 
The acronyms and the chemical compositions are explained in his paper. There is a good correlation between $T_{g}/\Delta T_{g}$ and $Q_{a}^{*}/k_{\rm B} T_{g}$.

From the approximation of Eq.~(\ref{eq:Approx-Qa}), Eq.~(\ref{eq:MoynihanFormula}) is rewritten as
\begin{equation}
\frac{E_{b,g}-E_{b,l}}{k_{\rm B} T_{g}} = 4.8.
\label{eq:EbtoTg} 
\end{equation}
This ratio seems to be reasonable when compared with the range of energy barrier $E_{b}$ of impurity diffusion. In silica glass, the energy barriers $E_{d}$ of impurity diffusions are reported to range from 0.3 to 0.8 eV \cite{Barrer34}. Since for silica glass $T_{g}=1450$ K, the ratio $E_{d}/k_{\rm B}T_{g} $ ranges from 2 to 6. Although diffusion and viscose motions are different modes of motions, the energy barriers should not be so different \cite{Avramov09,Brillo11}. Thus, the relationship shown in Eq.~(\ref{eq:EbtoTg}) must give reasonable values for $E_{b}$. This infers that the previous values for the activation energy for glass transition were significant overestimations.

Among various methods available, one definition of fragility is $m = (\partial \ln \eta/ \partial (1/T))/ k_{\rm B}T_{g}$, meaning the activation energy normalized with the glass-transition temperature \cite{Bohmer93}. Hence the relationship Eq.~(\ref{eq:MoynihanFormula}) is also rewritten as follows:
% how the relation $\eta$ vs.~$1/T$ deviates from the Arrhenius law. 
\begin{equation}
% \frac{Q_{a}^{*}}{k_{\rm B}T_{g} }  
m = 4.8 k.
\label{eq:Modify-Moynihan} 
\end{equation}
Although the factor was determined as 4.8 from a certain group of glasses, this may not largely change among various glasses: Ito {\it et al.}~showed that this is indeed the case \cite{Ito99}.
Thus, one can immediately see that fragility represents the magnification factor $k$. Therefore, the greater fragility, the more magnified $Q_{a}^{*}$ is compared to the true value $E_{b}$. Therefore, fragile glasses exhibit large $Q_{a}^{*}$, in spite of their low melting temperatures.
This naturally leads to the following interpretation of fragility $m$: that is, it represents how rapidly the energy barrier varies with varying $T$. This interpretation is more appealing because the energy barrier is the standard terminology in solid state physics and chemistry.
A similar interpretation for the $m$ was inferred by Dyre and Olsen \cite{Dyre04}. The present study gives an additional meaning of $m$ as the magnification factor for the activation energy
% however, the energy barrier in their relationships was the apparent activation energy $Q_{a}^{*}$.
Researchers attempted to find correlations between fragility with other properties of glasses. Fujimori and Oguni found the correlation of $m$ with a special index representing the difference between $\alpha$ and $\beta$ relaxations \cite{Fujimori95}. Scopigno {\it at al.}~found another correlation of $m$ with the decorrelation of the density fluctuations \cite{Scopigno03}. These correlations can be understood better by using a familiar term, namely, energy barrier.
% Schug98,Gramatp02,

%%%%%%%%%%%%%%%%%%%%%%%%%%%%%%%%%%%%%%%%%%%%%%
%%%%%%%%%%%%%%%%%%%%%%%%%%%%%%%%%%%%%%%%%%%%%%
%%%%%%%%%%%%%%%%%%%%%%%%%%%%%%%%%%%%%%%%%%%%%%
\section{Analysis of the temperature dependence of viscosity}
\label{sec:TdependViscosity}

\subsection{Silicate glass}
\label{sec:silicate}
Let us examine the full-scale temperature dependence of viscosity for three classes of glasses.
The first class of glasses comprises silicate glasses, which are typically strong glasses. It is said that the $T$ dependence of $\eta$ for strong glasses obeys the Arrhenius law, but it is true only in a relative sense compared with fragile glasses. If the $T$ dependence of $\eta$ is examined over a wide range of $T$, one sees a large variation in the activation energy: for example, the $Q_{a}^{*}$ of silica glass increases from 4.0 eV above $T=2000^{\circ}$C to 8.1 eV at $T=1400^{\circ}$C as temperature declines \cite{Mackenzie61,Hetherington64,Fontana68,Laughlin72}. 

\begin{figure}[htbp]
\centering
\includegraphics[width=140 mm, bb=0 0 441 240]{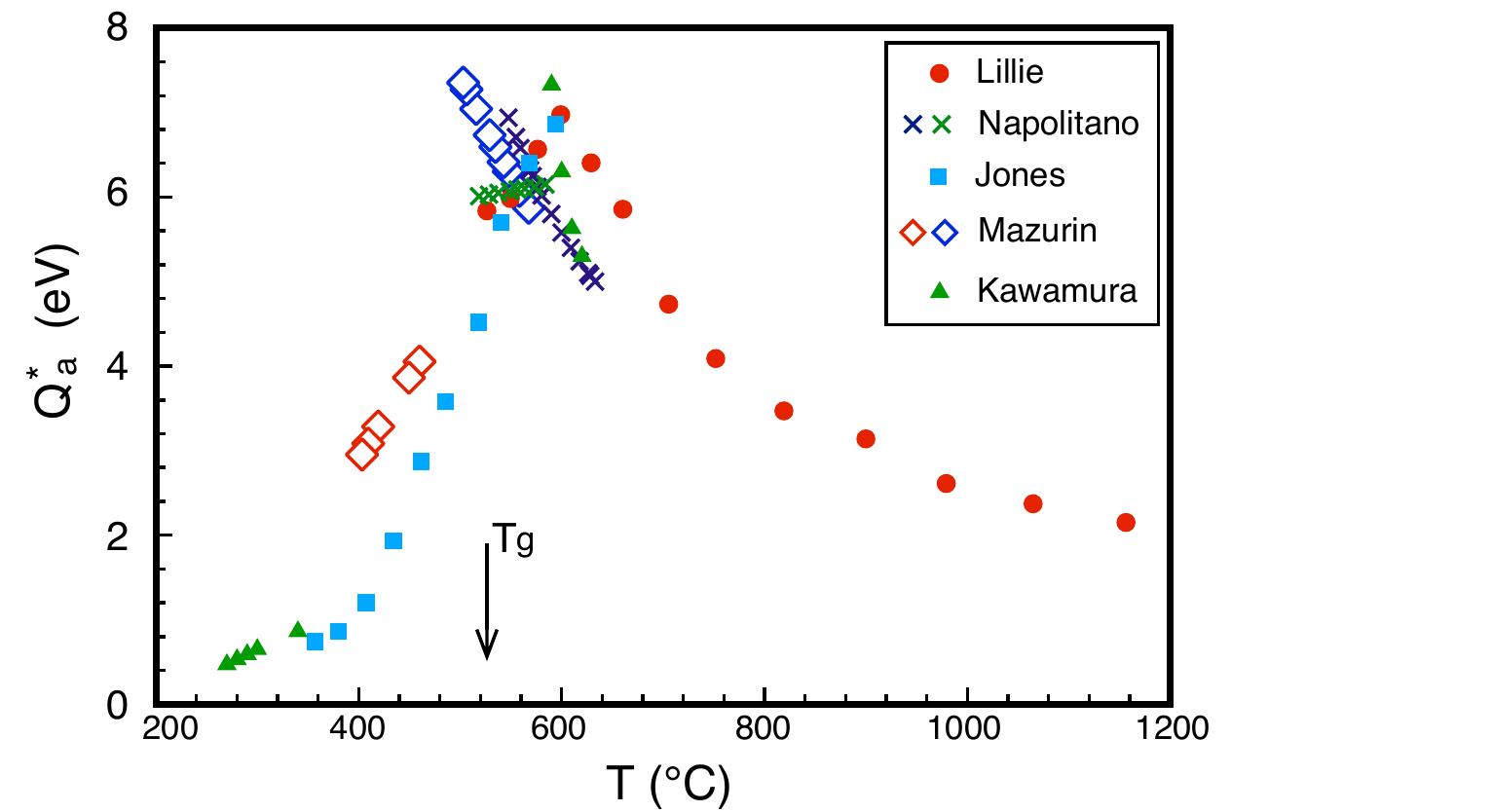} 
\caption{Apparent activation energy $Q_{a}^{*}$ of silicate glasses obtained by the conventional method.
Data are for soda-lime-silicate (SLS) glass samples, except for Kawamura's data. Kawamura's samples are high-level radioactive waste (HLW) glass.
The sources of data are as follows: Lillie \cite{Lillie31}, Jones \cite{Jones44}, Napolitano \cite{Napolitano64}, Mazurin \cite{Mazurin75}, and Kawamura \cite{Kawamura14}.
% For Napolitano, to kinds of samples (blue: stabilized fibers, green: annealed fibers) were measured \cite{Napolitano64}.
} \label{fig:all-Q}
\end{figure}

Figure \ref{fig:all-Q} shows the apparent activation energies $Q_{a}^{*}$ obtained using Eq.~(\ref{eq:QinArrhenius}) for soda-lime-silicate (SLS) glass, as reported by several authors. The chemical composition of SLS glass is formally ${\rm SiO_{2} : Na_{2}O : CaO = 75 : 15: 10}$, and the transition temperature $T_{g}$ is $530^{\circ}$C.
The numerical data of $\eta(T)$ were retrieved from the original figures by using a digitizer. The retrieved data were then processed by fitting them with smoothing functions, and $Q_{a}^{*}$ was obtained by taking the derivatives of the fitting function according to Eq.~(\ref{eq:QinArrhenius}). All the original data and the subsequent process of analysis are provided in Supplemental material.
As seen in Fig.~\ref{fig:all-Q}, the apparent activation energy $Q_{a}^{*}$ increases as temperature decreases to $T_{g}$. As the temperature approaches $T_{g}$, $Q_{a}^{*}$ reaches about 6.8 eV and then quickly decreases. The maximum value is more than twice the value at $T=1200^{\circ}$C. The linearity in the Arrhenius plot holds only for the range of $11 \le \log \eta \le 14$, which corresponds to a temperature range of $600 ^{\circ}$C $\ge T \ge 500^{\circ}$C.

\begin{figure}[htbp]
\centering
\includegraphics[width=140 mm, bb=0 0 441 240]{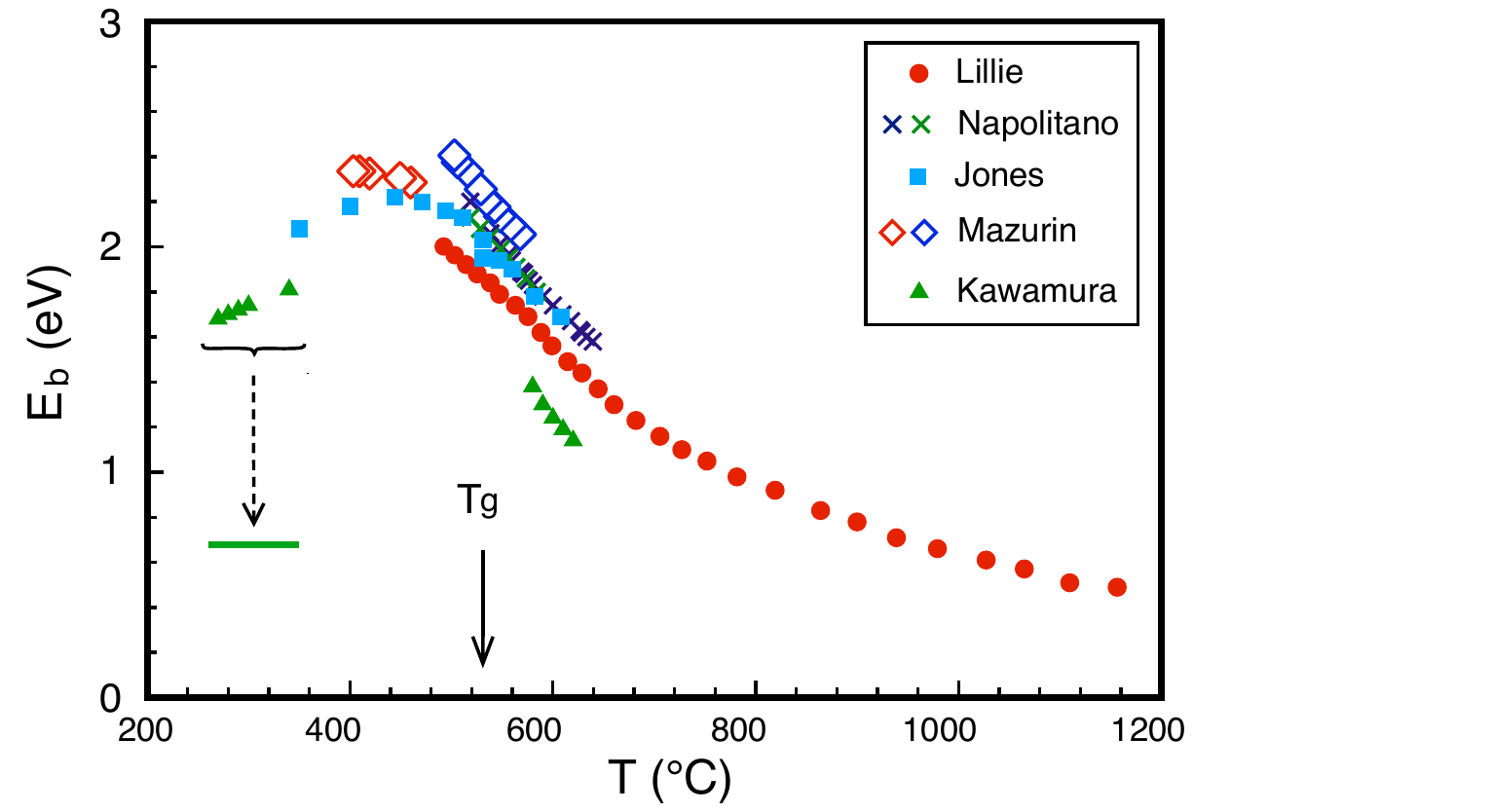} 
\caption{Energy barrier $E_{b}$ of SLS glass. The same symbols as those used in Fig.~\ref{fig:all-Q} are also used here. All data are calculated by assuming $\log \eta_{0}=0$. Under the assumption that $\log \eta_{0}=10.4$, the corrected values for Kawamura's data are reduced to the green line.
} \label{fig:true-Eb}
\end{figure}
Let us calculate the corrected value $E_{b}$ according to Eq.~(\ref{eq:trueEb}).
The constant $\eta_{0}$ is not known. % Nemilov wrote that $\log \eta_{0} = -3.5$.
The lowest value of the measured viscosity cited above is about $\log \eta = 2$, which was reported by Lullie \cite{Lillie31}.
The pre-exponential factor $\eta_{0}$ is less than this value but is larger than a typical value of normal liquids, i.e., $-3.5$. Since silicate glasses are among the most viscous materials, tentatively, $\log \eta_{0} = 0$ was used. 
The calculated values are plotted in Fig.~\ref{fig:true-Eb}.
The maximum value of $E_{b}$ reduces significantly to 2.2 eV from the corresponding value $Q_{a}^{*}$ of 6.8 eV. This $E_{b}$ value of 2.2 eV is already smaller than the Si-O bond energy of SiO$_{2}$ (4.6 eV). If an increase in $\eta_{0}$ with decreasing $T$ is taken into account, the value $E_{b}$ is further reduced. 
% According to Eq.~(\ref{eq:EbtoTg}), $E_{b,g} -E_{b,l} = 0.34$ eV.
%
The $E_{b}$ value can also be estimated by using Eq.~(\ref{eq:Approx-Qa}).
For the transition width $\Delta T_{g}$ of the SLS glass, the value of NBS710 ($T_{g}/\Delta T_{g} = 17$, as shown in Fig.~\ref{fig:Moynihan-plot}) is adopted. Equation (\ref{eq:Approx-Qa}) leads to $E_{b,g} - E_{b,l} = 0.37$ eV. The model shown in Eq.~(\ref{eq:EbdepT}) assumes that the change in $E_{b}$ occurs only within $\Delta T_{g}$. However, as seen in Fig.~\ref{fig:true-Eb}, the change in $E_{b}$ begins from far higher temperatures and is more gradual against $T$. In this respect, Eq.~(\ref{eq:Approx-Qa}) may be an overcorrection. The true value $E_{b}$ lies between 0.37 and 2.2, which covers the diffusion barrier $E_{d}$ of silica glass ($0.3 \sim 0.8 $ eV) mentioned above.

A notable observation in Fig.~\ref{fig:true-Eb} is the temperature dependence of $E_{b}$ below $T_{g}$. It is not easy to measure viscosity values larger than $\log \eta = 13$. Despite this difficulty, viscosity below $T_{g}$ were measured even in early studies \cite{Stanworth37,Jones44,Jones49}. 
For SLS glass, Jones reported a significant reduction in $Q_{a}^{*}$ from 2.3 eV above $T_{g}$ to 0.25 eV at $T=350^{\circ}$C, at which $\eta$ increases to about $\log \eta = 18$ \cite{Jones44,Jones49}. The values reported by Jones are shown in Fig.~\ref{fig:all-Q}: these values are obtained by the present analysis employing a smoothing functions, and hence the values are slightly different from the values reported by Jones. A similar reduction was reported by Shen {\it et al} \cite{Shen03}: a reduction from $Q_{a}^{*}=$5.2 eV above $T_{g}$ to 1.2 eV at $T=450^{\circ}$C was observed for SLS glass.
Surprisingly, on applying the correct form of Eq.~(\ref{eq:trueEb}), these quick reductions in $Q_{a}^{*}$ below $T_{g}$ are drastically modified so that they are nearly constant, as shown in Fig.~\ref{fig:true-Eb}.
This is consistent with the present model shown in Fig.~\ref{fig:Glass-trans}(c). This behavior is reasonable because the structure does not change once the glass substance is frozen. There are some literatures describing the activation energy as being saturated around $T_{g}$ \cite{Mazurin82,Nemilov-VitreousState}, which means a recovery of the Arrhenius law below $T_{g}$. 
Recently, Kawamura {\it et al.}~performed viscosity measurements at temperatures as low as $270^{\circ}$C, which may be the lowest temperature ever reported for silicate glasses \cite{Kawamura14}. They employed the fiber-bending method, which was developed by Koide \cite{Koide94}. They performed measurements for a simulated high-level radioactive waste (HLW) glass. The main difference between HLW and SLS glass is inclusion of B$_{2}$O$_{3}$ in HLW.  Hence, it is reasonable to observe that the viscosity of HLW glass was lower than that of SLS glass.
If, for Kawamura's data, $\log \eta_{0}$ is calculated to make $\eta(T)$ constant below $T_{g}$, the pre-exponential factor is obtained as $\log \eta_{0}=10.4$, as indicated by the green line in Fig.~\ref{fig:true-Eb}.

All the presented data were measured by independent authors and by different methods. Therefore, the agreement between the values of $E_{b}$ below as well as above $T_{g}$ validates the reliability of the obtained $T$ dependence. The energy barrier $E_{b}$ becomes saturated below $T_{g}$.
By taking all the above results together, it is likely that $E_{b}$ of SLS glass is no more than 1 eV.
Furthermore, the present analysis shows that no divergence occurs at $T_{0}$. This conclusion agrees with the conclusions of recent studies on the glass transition \cite{Hecksher08,Zhao13,Ponga15}.

\subsection{Organic glass}

\begin{figure}[htbp]
\centering
\includegraphics[width=140 mm, bb=0 0 364 240]{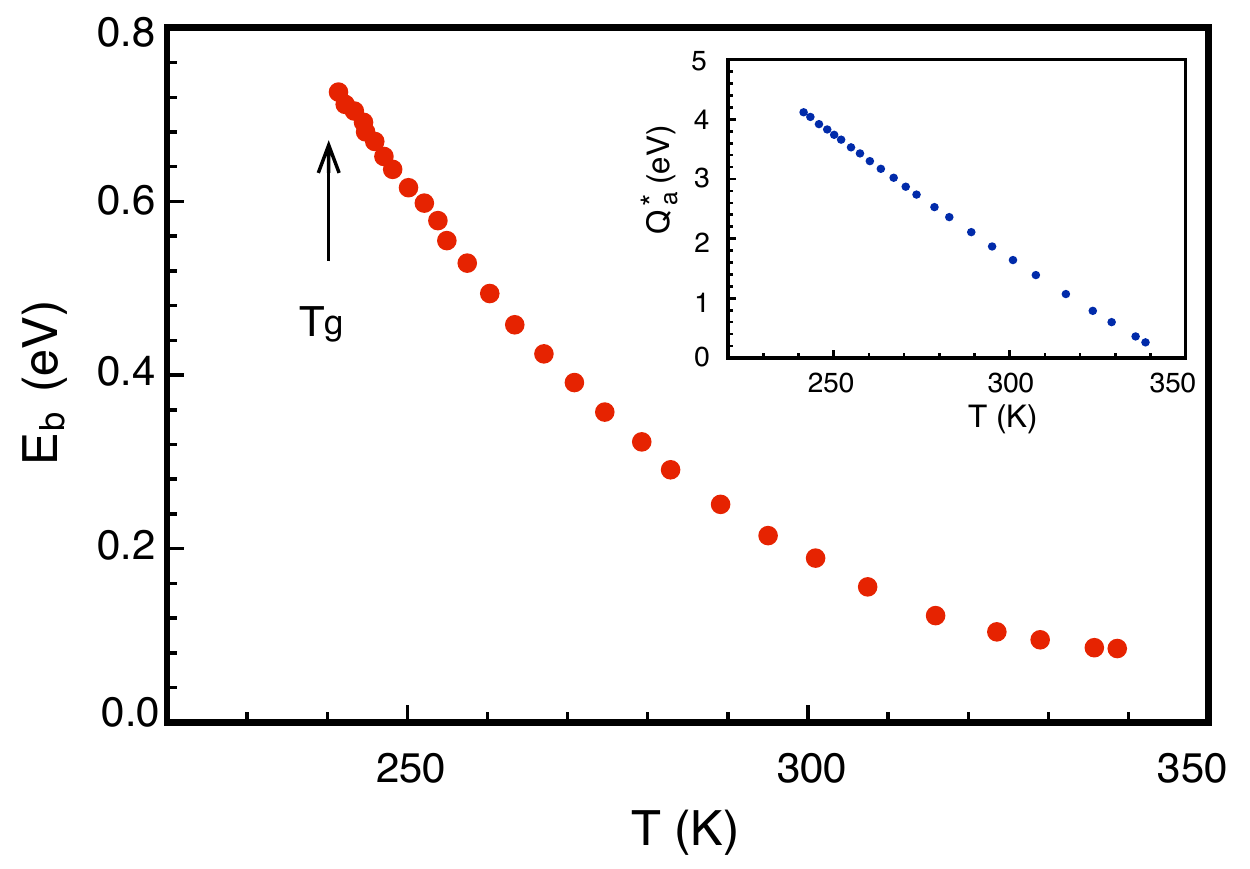} 
\caption{Energy barrier $E_{b}$ of $o$-terphenyl glass. $T_{g}=239$K. The inset shows the apparent activation energy $Q_{a}^{*}$ obtained using the traditional Arrhenius plot. The data are taken from Ref.~\cite{Nemilov-VitreousState}.
} \label{fig:Terphenyl_Eb}
\end{figure}

The next class of glasses to be examined comprises organic glasses. Generally, the glasses belonging to this class have low-melting temperatures that are near or lower than room temperature. Most of them are fragile glasses, and hence the apparent activation energy $Q_{a}^{*}$ strongly depends on $T$. When $T$ approaches to $T_{g}$ from high temperatures, $Q_{a}^{*}$ increases quickly, as shown in the Angell plot. Large values of $Q_{a}^{*}$ are reported at $T_{g}$: e.g., $Q_{a}^{*}$ = 5.5 eV for glucose and 1.1 eV for glycerol \cite{Davies53}. More examples are provided in the review by Hodge, who has presented many values of $Q_{a}^{*}$ in different ways depending on the definition of the activation energy: apparent activation energies exceeding 10 eV are listed \cite{Hodge94}.
In this work, $o$-terphenyl is examined as an example of organic glasses. This is because the viscosity of this material was measured in a wide range from $\log \eta = -0.5$ to $13$, which covers the whole range of the Angell plot. Glass transition occurs at $T_{g}=239$K with the width $\Delta T_{g}=5$K. For model calculations of the $T$ dependence of $o$-terphenyl, see Ref.~\cite{Nemilov-VitreousState}. The inset of Fig.~\ref{fig:Terphenyl_Eb} shows $Q_{a}^{*}$ of $o$-terphenyl as obtained by the conventional method using Eq.~(\ref{eq:QinArrhenius}). 
% (0.02 eV) 
The data $\eta(T)$ were retrieved from Nemilov's textbook (Fig.~44 of Ref.~\cite{Nemilov-VitreousState}): the plot data actually consist of three sets of data provided by Laughlin \cite{Laughlin72}, Greet \cite{Greet67}, and Cukierman \cite{Cukierman73}.
The apparent activation energy $Q_{a}^{*}$ increases up to 4 eV, which is close to that of silicate glass. 
In the literature, this large $Q_{a}^{*}$ for organic glasses is at times explained by the collective modes of a large moving units. This explanation is not convincing. A collective mode can alter $\lambda$ in the pre-exponential factor in Eq.~(\ref{eq:A-factor}). However, each atom constituting the moving unit receives equally thermal energy $k_{\rm B}T$. Hence, it is unlikely that the activation energy increases with the increase in the size of moving units. 
% \cite{Leibfried57}.
% It is unrealistic for a low-melting-point glass to have an energy barrier as large as that of high-melting point glass.
This unreasonably large value can, however, be corrected by the present theory. By applying Eq.~(\ref{eq:trueEb}), a significant reduction from $Q_{a}^{*}$ is obtained. This is shown in Fig.~\ref{fig:Terphenyl_Eb}. In this case, $\log \eta_{0} = -2$ is used. The maximum $E_{b}$ is 0.8 eV, which is within a reasonable range of the energy barrier. If the multiplication factor, in this case $k=50$, is adopted, the value $E_{b}$ becomes as small as 0.1 eV, when $E_{b,l}$ is ignored.
Now, it is clear that the large values of the apparent activation energy $Q_{a}^{*}$ that are common for organic glasses can be ascribed to the narrow width $\Delta T_{g}$.

\subsection{Metallic glass}
The third example comprises metallic glasses. Metallic glasses have fragility intermediate between that of strong covalent glasses and fragile organic glasses \cite{Busch07}. The apparent activation energy $Q_{a}^{*}$ of the glass transition of metallic glasses ranges from 2 to 8 eV near $T_{g}$ \cite{Johnson07}. 
The plastic deformation of crystalline materials is mediated by dislocations, whereas for metallic glasses it is madiated by the so-called shear transformation zone (STZ) \cite{Argon77}. 
The approach for describing the $T$ dependence of the viscosity is opposite to that employed in Sec.~\ref{sec:TheoryOfViscosity}, where the starting state is the liquid state. For the area of metallic glasses, the study starts from the solid state \cite{Schuh07}.
The formula for viscosity is formally the same as that given in Eq.~(\ref{eq:Act-eta}). % while the physical meaning of the pre-exponential factor $\eta$ is different. 
The activation volume $V_{a}$ is considered to be that of an STZ. For ZrTiCuNi glass, a typical STZ includes 20--30 atoms \cite{Schuh07}. % This gives the activation energy of the size $Q_{a} = 4.6$ eV.

The temperature dependence of the viscosity has been gleaned from recent experiments on metallic glasses.
A theoretical model was proposed by Johnson {\it et al.}, who considered a formula for the energy barrier based on the elastic moduli of the solid state and extended it to the liquid state \cite{Demetriou06,Johnson07}. 
They adapted the exponential decay for the $T$ dependence of the energy barrier, obtaining a formula 
\begin{equation}
\frac{\eta}{\eta_{0}}=\exp \left\{ \frac{W_{g}}{k_{\rm B}T} 
	\exp \left[ 2n \left( 1- \frac{T}{T_{g}} \right) \right] \right\},
\label{eq:Johnson} 
\end{equation}
for $T > T_{g}$. Here, $W_{g}$ is the energy barrier at $T=T_{g}$, and $n$ is a fitting parameter whose value is the order of unity \cite{Johnson07}. 
% When $T$ is close to $T_{g}$, this formula becomes the Arrhenius form, giving the apparent activation energy being $W_{g}$. 
% The values $W_{g}$, which were obtained by fitting to experimental data, are typically 1 to 2 eV. In this case, values of $Q_{a}^{*}$ are not too large.
When $T$ approaches $T_{g}$, the term $\exp \left[ 2n (1-T/T_{}g) \right] $ in Eq.~(\ref{eq:Johnson}) can be expanded with respect to $\Delta T = T-T_{g}$, resulting in
\begin{equation}
\frac{\eta}{\eta_{0}}=\exp \left\{ \frac{W_{g}}{k_{\rm B}T} 
	\left( 1+2n -2n \frac{T}{T_{g}}  \right)  \right\}.
\label{eq:Johnson1} 
\end{equation}
This has the same form as the linear approximation of Eq.~(\ref{eq:EbdepT}) with a magnification factor $k=1+2n$.
Numeric examples are given by them: for ${\rm Pd_{77.5}Cu_{6}Si_{16.5}}$, $T_{g}=634$K, $\log \eta_{0}=-2.11$, and $n=1.67$. This leads to the magnification factor $k$ of $4.3$. 
Hence, the correction for $E_{b}$ is not too large compared with those of other classes of glasses.

An interesting point regarding metallic glasses is that viscosity below $T_{g}$ is sometimes measured because of the practical interest in the creep phenomenon. Taub and Spaepen found that the activation energy $Q_{a}^{*}$ of viscosity of ${\rm Pd_{82}Si_{18}}$ is almost constant at about 2.0 eV in a temperature range from 420 to 540 K \cite{Taub80}; ${\rm Pd_{82}Si_{18}}$ is a metallic glass with $T_{g}=634$ K \cite{Wang17}. Interestingly, the value  $Q_{a}^{*}$ does not change by annealing at various temperatures. This is consistent with the present assumption, Eq.~(\ref{eq:EbdepT}).
There is no sign to exhibit the divergent behavior of the VFT law in this case too.
% Thus, it is inferred that $T_{0}$ is merely a fitting parameter to reproduce the $T$ dependence of $\eta$ above $T_{g}$.

% Taub79,

\section{Discussion} 
\label{sec:discussion}
Although phase transition was not the original focus of this study, the findings of this study have further implication to this issue. The analyses in Sec.~\ref{sec:TdependViscosity} established that none of the experimental data on viscosity below $T_{g}$ showed any sign of exponential divergence at a finite temperature. The $T$ dependence of $\eta(T)$ seems to obey the Arrhenius law rather well.
Indirect evidence for this non divergence was recently reported \cite{Hecksher08,Zhao13,Ponga15}.
% Often repeated claim is that if we wait for an extremely long time, it would diverge is unrealistic.

Both $T_{0}$ and $T_{K}$ are obtained by extrapolating viscosity and entropy, respectively, measured at $T>T_{g}$. This extrapolation loses its validity when one accepts the experimental transition occurring at $T_{g}$ as a genuine phase transition. 
% Although the transition is not so abrupt as that of crystal, the broadening of the transition temperature is merely a consequence of the lack of long-range order. 
Nobody considers that an extrapolation of the $T$ dependence of entropy ($S \sim \ln T$) of an ideal gas to $T=0$ is meaningful: such extrapolation diverges negatively.
The repeated assertion is that the observed glass transition occurring at $T_{g}$ is not an intrinsic property of a glass because $T_{g}$ is affected by the preparation conditions. This subject was already discussed in Ref.~\cite{Shirai20-GlassState}. The essential points are repeated herein.
First, the kinetic nature of the transition is common to all liquid/solid phase transitions. Crystallization is determined by the competition of the kinetic effect of entropy and the potential effect on constraining atomic motions.
Second, preparation dependence is common in every transition, though the effect is outstanding for glasses. Freezing of water to ice is a well-known example. The graphite-diamond transition temperature is largely affected by its kinetics. Any crystallization accompanies a kinetic process of nucleation that can be altered by external conditions.
The reference temperature $T_{0}$ in the VFT formula is only a fitting parameter to reproduce the temperature dependence of viscosity above $T_{g}$.

%%%%%%%%%%%%%%%%%%%%%%%%%%%%%%%%%%%%%%%%%%%%%%%%%
\section{Conclusion} 
\label{sec:conclusion}
In usual phase transitions, the energy barrier for atomic movement changes abruptly. In fact, the transition occurs at a fixed temperature, such as $T_{m}$, which hides this change from observation. In contrast, for glass transition, the continuous structural change occurs in a certain range of temperature, i.e., $\Delta T_{g}$.
This structural change causes a change in the energy barrier $E_{b}$ in a continuous manner, and this causes $\eta(T)$ to deviate from the Arrhenius law.
The present theory predicts that the observed value of the activation energy $Q_{a}^{*}$ largely overestimates the energy barrier $E_{b}$. This prediction has been validated by examining available experimental data.
The degree of deviation depends mainly on the magnification factor $k=T_{g}/\Delta T_{g}$.
Generally, fragile glasses exhibit sharp transitions with narrow widths $\Delta T_{g}$. This explains the extraordinarily large values for the observed $Q_{a}^{*}$ of fragile glasses: e.g.~the magnification factor is on the order of magnitude 50. This magnification also occurs in strong glasses, whereas the magnification factor is not very large. In either case, the genuine energy barrier $E_{b}$ is estimated to be less than 1 eV, which lies in a reasonable range of the chemical energy of liquids and solids.
The analysis has provided another interpretation for fragility of glass: fragility indicates how fast the energy barrier varies with varying temperature.
Thus, the present result enables us to interpret the behavior of the viscosity of glass on the same ground of energetic approach that is commonly used in solid-state physics and chemistry. The results also provide a starting point of future study by first-principles calculations.

A shortcoming of the current status of this theory is lack of the concrete formula for the pre-exponential factor $\eta_{0}$. Since the present assumption of Eq.~(\ref{eq:EbdepT})---the energy barrier changes only within $T_{g}$---is not very good, the lack of the concrete formula for $\eta_{0}$ can cause a serious problem. Experimental data given in Sec.~\ref{sec:TdependViscosity} show that the energy barrier begins to change at temperatures far higher than $T_{g}$. Clearly, this shortcoming should be solved.

% \label{sec:conclusion}

%%%%%%%%%%%%%%%%%%%%%%%%%%%%%%%%%%%%%%%%%%%%%%%
\begin{acknowledgments}
% The author thanks P. D. Gujrati for valuable debates on glass physics and Yamamuro for helpful discussion on the specific heat of glasses.
This work was performed under the Research Program of ``Dynamic Alliance for Open Innovation Bridging Human, Environment and Materials" in ``Network Joint Research Center for Materials and Devices".
The author also thanks Enago (www.enago.jp) for the English language review.
%The author received a financial support by the Research Program of ``Five-star Alliance" in ``NJRC Mater.~\& Dev."
\end{acknowledgments}

%%%%%%%%%%%%%%%%%%%%%%%%%%%%%%%%%%%%%%%%%%%%%%%

%%%%%%%%%%%%%%%%%%%%%%%%%%%%%%%%%%%%%%%%%%
% \bibliography{thermo-refs,glass-refs}

%%%%%%%%%%%%%%%%%%%%%%%%%%%%%%%%%%%%%%%%%%
\end{document}